# Importance of the crystalline symmetry in the piezoelectric properties of $(K_{0.44+x}Na_{0.52}Li_{0.04})(Nb_{0.86}Ta_{0.10}Sb_{0.04})O_{3+x/2}$ lead-free ceramics.

F. Rubio-Marcos[1,*], P. Marchet[2], T.Merle-Méjean[2], J.F. Fernandez[1].

[1] *Electroceramic Department, Instituto de Cerámica y Vidrio, CSIC, Kelsen 5, 28049 Madrid, Spain.*
[2] *SPCTS, UMR 6638 CNRS, Université de Limoges, 123, Av. A. Thomas, 87060 Limoges, France.*

Lead-free ceramics $(K_{0.44+x}Na_{0.52}Li_{0.04})(Nb_{0.86}Ta_{0.10}Sb_{0.04})O_{3+x/2}$ (x=-0,06) were prepared by conventional solid state sintering. The results indicate a correlation between crystalline symmetry and electrical properties. Higher tetragonality was observed in samples with higher piezoelectric response. As the tetragonality increase the presence of polymorphic phase decrease associated with a reduction of the relaxor behaviour response. The ceramics with c/a = 1.011 ratio exhibit enhanced electrical properties, $d_{33}$~230 pC/N; $K_p$~0.42 %; $K_t$~0.36 %; $T_c$~270ºC. The Raman spectroscopy showed a certain asymmetry in the paraelectric phase above the Curie temperature $T_c$ that provokes retention of the polarization. The crystalline symmetry plays a crucial role in the piezoelectric properties of the system.

***Corresponding author.**

**Electronic address: frmarcos@icv.csic.es**

The search for alternative lead-free piezoelectric materials is now being focused on modified bismuth titanates,[1] alkali niobates and systems in which a morphotropic phase boundary (MPB) occurs.[2,3,4] Over the past few years, much attention for lead-free piezoelectric ceramics has been given to $(Na_{0.5}K_{0.5})NbO_3$ (KNN) based piezoelectric ceramics because of their good electrical properties.[5] KNN exhibits a MPB and as for lead titanate zirconate piezoceramics, PZT, an increase in properties for compositions close to this MPB. However, the major drawbacks of KNN ceramics are (i) the need for special handling of the starting powders due to volatility of alkaline elements, (ii) high sensitivity of the properties to stoichiometry, and (iii) complex densification processes.[6]

Excess of alkaline elements in the ceramic samples easily reacts with the moisture in the air and shows deliquescence.[7] Moreover alkaline metal elements included in these materials easily evaporate at high temperatures. In addition, the evaporation of one of the constituents of the MPB provokes thus a compositional fluctuation that results in poorer properties, as it was well known for the piezoelectric PZT perovskite system.[8] In order to solve the problem of densification, different advanced processes such as spark plasma sintering[9] or hot pressing,[10] were used to promote the sintering which led to the enhanced density and properties. However, their cost is high.

The sinterability of these materials should be improved by sintering aids as CuO[6,9,11] because its low melting point. The sintering aids usually enter in B position of the $ABO_3$ perovskite structure and thus the A site vacancies suppress the formation of the hygroscopic secondary products.[12]

Recently it was reported exceptionally high piezoelectric properties in the system $(K,Na)NbO_3$-$LiTaO_3$-$LiSbO_3$[13] prepared by a complex processing method, with $d_{33}$ values over 400 pC/N for textured ceramics. In this system the excess of B-perovskite cations improved the densification[6] of the samples processed by conventional routes. Actually, there are some structural and electrical aspects that remain controversial because of novelty of the system. In the case of the Curie temperature (Tc) values, Saito *et al*[13] reported values of 252ºC while other authors consider it around 337ºC[14]. It was also established that tetragonal compositions presented higher piezoelectric coefficients ($d_{ij}$), but same nominal compositions reached different Tc and $d_{33}$ values. Another aspect that remains controversial is the MPB nature of the system that in some cases was reported to be a polymorphic behaviour[15,16] and shown compositional inhomogeneities.[17] Therefore the aim of this work is to establish a relationship between the crystalline symmetry and the dielectric and piezoelectric properties as a function of the sintering time of this system.

Since in KNN system the MPB is correlated to the Na/K ratio, starting from the *Saito et al* composition[13], several samples of global formula $(K_{0.44+x}Na_{0.52}Li_{0.04})(Nb_{0.86}Ta_{0.10}Sb_{0.04})O_{3+x/2}$, corresponding to different potassium content were studied (the corresponding results will be published elsewhere). The composition $(K_{0.44+x}Na_{0.52}Li_{0.04})(Nb_{0.86}Ta_{0.10}Sb_{0.04})O_{3+x/2}$ (x=-0,06) was selected to be of tetragonal symmetry and account for ratio of A/B perovskite cations of 0.94, in order to reduce the presence of orthorhombic phase. Hereafter, it will be named $K_{0.38}$NLNTS. The corresponding composition was prepared by a conventional ceramic processing route. $Na_2CO_3$ (99.5%), $Li_2CO_3$ (99.5%), $K_2CO_3$ (99%), $Nb_2O_5$ (99.9%), $Ta_2O_5$ (99 %), and $Sb_2O_5$ (99.995%) were used as starting raw materials. The method to optimize the particle size distribution of the raw materials was previously[18] described. The powders were attrition-milled using $ZrO_2$ balls in ethanol medium for 3h, dried, calcined at 700 °C for 2h, attrition milled again and pressed at 200 MPa into disks of 10 mm diameter and 0.7 mm thickness. The pellets were sintered at 1125 °C for different times ranging from 1h to 16h. For the electrical measurements a fired silver paste was used. The samples were poled in a silicon oil bath at 25 °C by applying a direct current electric field of 4.0 kV/mm for 30 min.

The crystalline symmetry was examined by x-ray diffraction analysis (XRD; Siemens Kristalloflex, CuKα radiation) using Si as internal standard. The lattice parameters were refined by a method of global simulation of the full diagram using the *fullprof* program. The Raman scattering was excited using 514 nm radiation from an $Ar^+$ laser operating at 10 mW and it was collected by a microscopic Raman spectrometer (Jobin Yvon 64000) in the 100 $cm^{-1}$-1100 $cm^{-1}$ range at room temperature. For temperature-dependent Raman measurements, a calibrated hot stage was used with a temperature stability of ±1 ºC. The temperature dependence of the dielectric constant of the ceramics was examined using a programmable furnace with an impedance analyzer (Agilent 4294A) in the frequency range 100Hz-1MHz. The piezoelectric constant $d_{33}$ was measured using a piezo-$d_{33}$ meter (APC YE2730A). The electromechanical coupling factors $k_p$ and $k_t$ were determined at room temperature from resonance and antiresonance methods on the basis of IEEE standards.

Figure 1(a) shows a selected part of the XRD patterns of ($K_{0.38}$NLNTS) ceramics as a function of the sintering time. The well crystallised samples show slight evolution of the tetragonal (002) and (200) diffraction peaks with the sintering time. The figure 1(b) presents the evolution of the intensity ratio of the two characteristic diffraction peaks $I_{(002)}/I_{(200)}$, that revealed a clear evolution of the system toward a pure tetragonal ratio of 0.50. This structural evolution accounts for the initial presence of a phase which can be assigned to the orthorhombic symmetry[16]. The only processing condition that varied in these samples was the sintering time. As a consequence, the possible origins of this phenomenon are (i) an evolution of the grain size (ii) an evolution of the composition and (iii) a chemical homogenization of the A and B perovskite ions, since the stoichiometry of these samples is very complex. The measured grain size barely changes from 1.36 ± 0.87 µm to 1.70 ± 1.04 µm, for the ceramics with 1h and 16h sintering time respectively and cannot explain such a structural change. An evolution of the chemical composition cannot be totally excluded, particularly for the alkaline elements, but, as we will see later (evolution of the dielectric properties), the most probable origin of this behaviour could be related to chemical homogenization with the sintering time as was reported in PZT system.[8]

In addition, the evolution of the crystalline symmetry is confirmed by raman spectroscopy. A displacement of the bands towards higher Raman shift values is observed with the increase of the sintering time, fig. 1(c). The main vibrations are associated to the $BO_6$ perovskite-octahedra.[19,20] In particular, $\upsilon_1$ and $\upsilon_5$ are detected as relatively strong scatterings in systems similar to the one we are studying because of a near-perfect equilateral octahedral symmetry. The $\upsilon_1$ peak, shift to higher frequency is due to an increase in strength constant force caused by the shortening of the distance between $B^{+5}$ type ions and their coordinated oxygens.[21]

The correlation of the crystalline structure of the samples with the piezoelectric properties was clearly evidenced in fig. 2. A similar increase of tetragonality ratio (c/a), piezoelectric constant $d_{33}$ and the wave number ($cm^{-1}$) for $\upsilon_1$ with the sintering time is clear from this figure. Therefore, this evolution reveals a relationship between the crystalline symmetry and the piezoelectric properties: the higher the tetragonality ratio of the system the higher the piezoelectric properties are.

The dielectric constants and dielectric losses (100 kHz) as a function of temperature for unpoled samples are shown in Fig. 3(a). The curves shows only presents at 270ºC a clear peak of the relative permittivity that corresponds to the ferroelectric–paraelectric phase transition. The Curie temperature slightly increases as a function of the sintering time but this increase was under the sensitivity of the $T_c$ measurement ~5ºC. However, the diffuseness of the phase transition can be determined from the modified Curie-Weiss law, $1/\varepsilon_r - 1/\varepsilon_m = C^{-1}(T-T_m)^{\gamma}$ [22] for which $\gamma = 2$ corresponds to a relaxor behaviour while $\gamma = 1$ corresponds to a classical ferroelectric–paraelectric phase transition.[23] Figure 3(b) shows the plots of $\ln(1/\varepsilon_r - 1/\varepsilon_m)$ vs $\ln(T-T_m)$ for the $K_{0.38}$NLNTS ceramics with sintering time between 1h and 16h. All the samples exhibit a linear relationship. The $\gamma$ value was determined by least-squares fitting of the experimental data to this modified Curie-Weiss law. For the ceramics with 1h sintering time, the $\gamma$ value calculated is 1.53, suggesting a relevant relaxor ferroelectric behaviour. As sintering time increases, $\gamma$ diminishes gradually, fig. 3(c), reaching a value of 1.36 for 16h sintered pellets, indicating that the ceramic has been evolved from a relaxor state toward a normal ferroelectric state. As discussed earlier, $\gamma$ values indicate that local disorder and clusters could be on the origin of this behaviour for $K_{0.38}$NLNTS materials and decreased with the sintering time as a result of solid state chemical homogenization. Such disorder and clusters were presented in PZT because solid solutions tend to have compositional fluctuation.[8,14,24]

The ferroelectric–paraelectric phase transition was assigned to tetragonal and cubic[25] crystalline symmetries, respectively. The occurrence of the phase transition is confirmed by Raman spectroscopy for the $K_{0.38}$NLNTS 16h sintering time sample, fig. 4(a).The Raman spectra as a function of the temperature was fitted using a sum of Lorenzian lines. Information on width, intensity, and wavenumber of the $\upsilon_1$ peak was extracted from the fitted curves and it is shown as a function of the temperature in figure 4(b). The full width at half maximum (FWHM) and the $\upsilon_1$ wavenumber ($cm^{-1}$) demonstrate a discontinuity at ~ 270ºC due to the phase transition. Generally for perovskite systems, the high temperature paraelectric phase has a cubic crystalline symmetry and does not support any Raman active modes at all. In the case of our samples and as for other perovskite compounds[20], Raman bands are observed above the phase-transition temperature, which means that the high-temperature symmetry is pseudo-cubic and probably presents a local order which differs from the global one. This certain asymmetry above $T_c$ could provoke retention of the polarization.

In conclusion, $K_{0.38}$NLNTS lead-free piezoelectric ceramics were developed by a conventional solid-state processing route. The effects of the sintering time on the crystalline symmetry and electrical properties of the ceramics were investigated. This relation between structure and properties indicate that this system evolves to a stable ferroelectric phase, with the sintering time. The initial presence of a polymorphic phase, coexistence of chemically disordered orthorhombic-phase in a tetragonal phase, seems to be responsible of partial relaxor behaviour since the system does not totally follow the Curie-Weiss law. The diminution of the orthorhombic phase percentage increases the proportion of the tetragonal phase improving the piezoelectric properties. In addition, the paraelectric phase, pseudocubic, showed a certain asymmetry above $T_c$ that could correspond to the presence of local non-cubic clusters. Therefore, in the design of these materials, it will be necessary to obtain structures with high tetragonality ratio.

This work was financially supported by the Spanish CICYT under contract MAT2007-66845-C02-01 and by a grant from FPI-CAM-FSE program.

**Figure Captions:**

**Figure 1. (color online)** (a) XRD patterns from 44º to 47º $2\theta$, of the $K_{0.38}$NLNTS ceramics (b) The intensity ratio of both diffraction peak $I_{(002)}/I_{(200)}$, (c) Raman spectra of the ceramics sintered at different times.

**Figure 2. (color online)** Variations of tetragonality ratio (c/a), $d_{33}$ and wave number ($cm^{-1}$) of $\upsilon_1$ stretching mode with the sintering time for the $K_{0.38}$NLNTL ceramics.

**Figure 3. (color online)** (a) Temperature dependence of $\varepsilon_r$ for the ceramics with different sintering times. (b) Plots of for the $K_{0.38}$NLNTL ceramics. The symbols denote experimental data, while the solid lines correspond to the least-squares fitting of the modified Curie-Weiss law. (c) Diffuseness, $\gamma$, of the ceramics as a function of the sintering time.

**Figure 4. (color online)** (a) Raman spectra as a function of the temperature for the 16h sintered sample. (b) Full Width at Half Maximum (FWHM) and wave number of $\upsilon_1$ stretching mode as a function of the temperature.

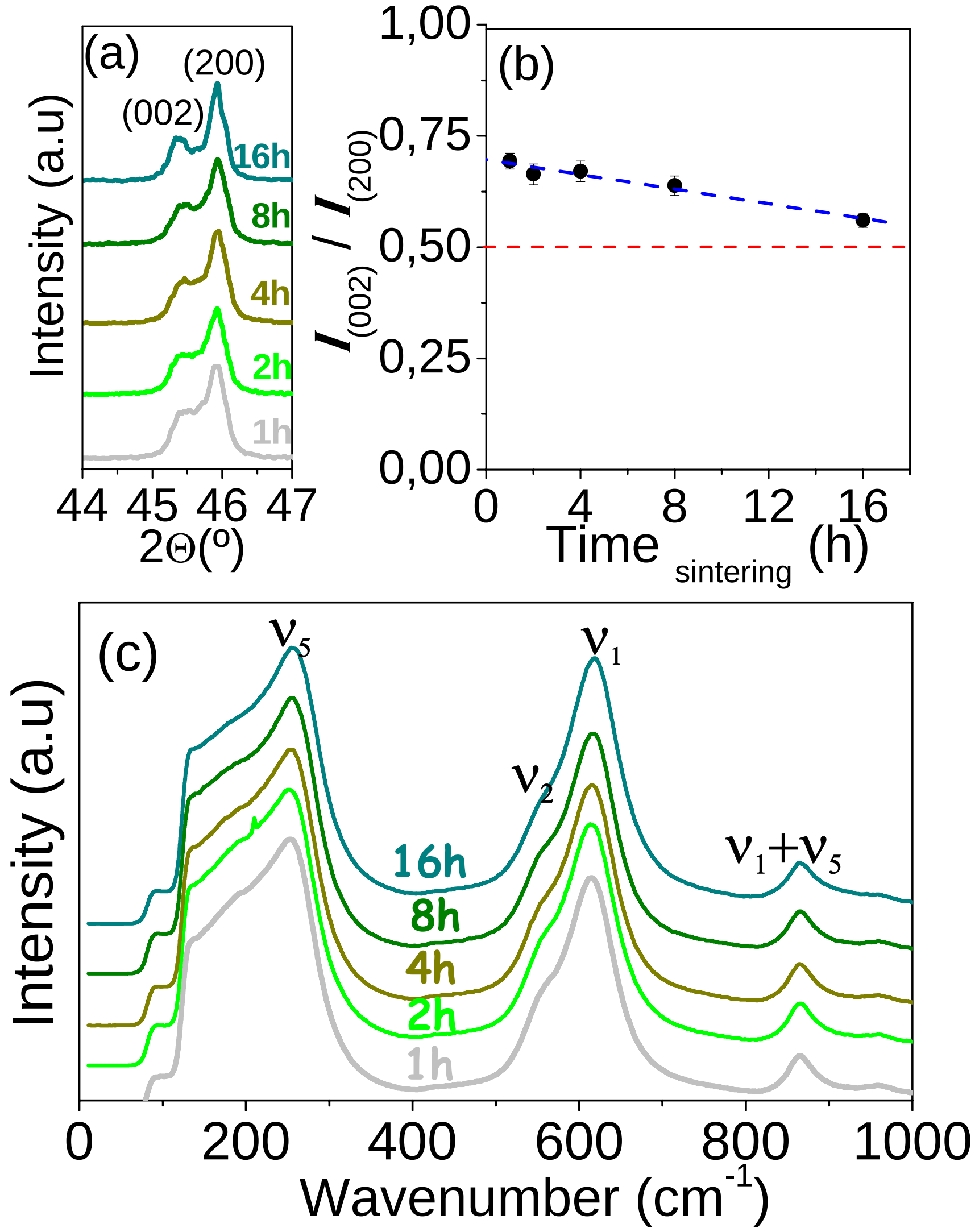

**F.Rubio-Marcos et.al. Figure. 1**

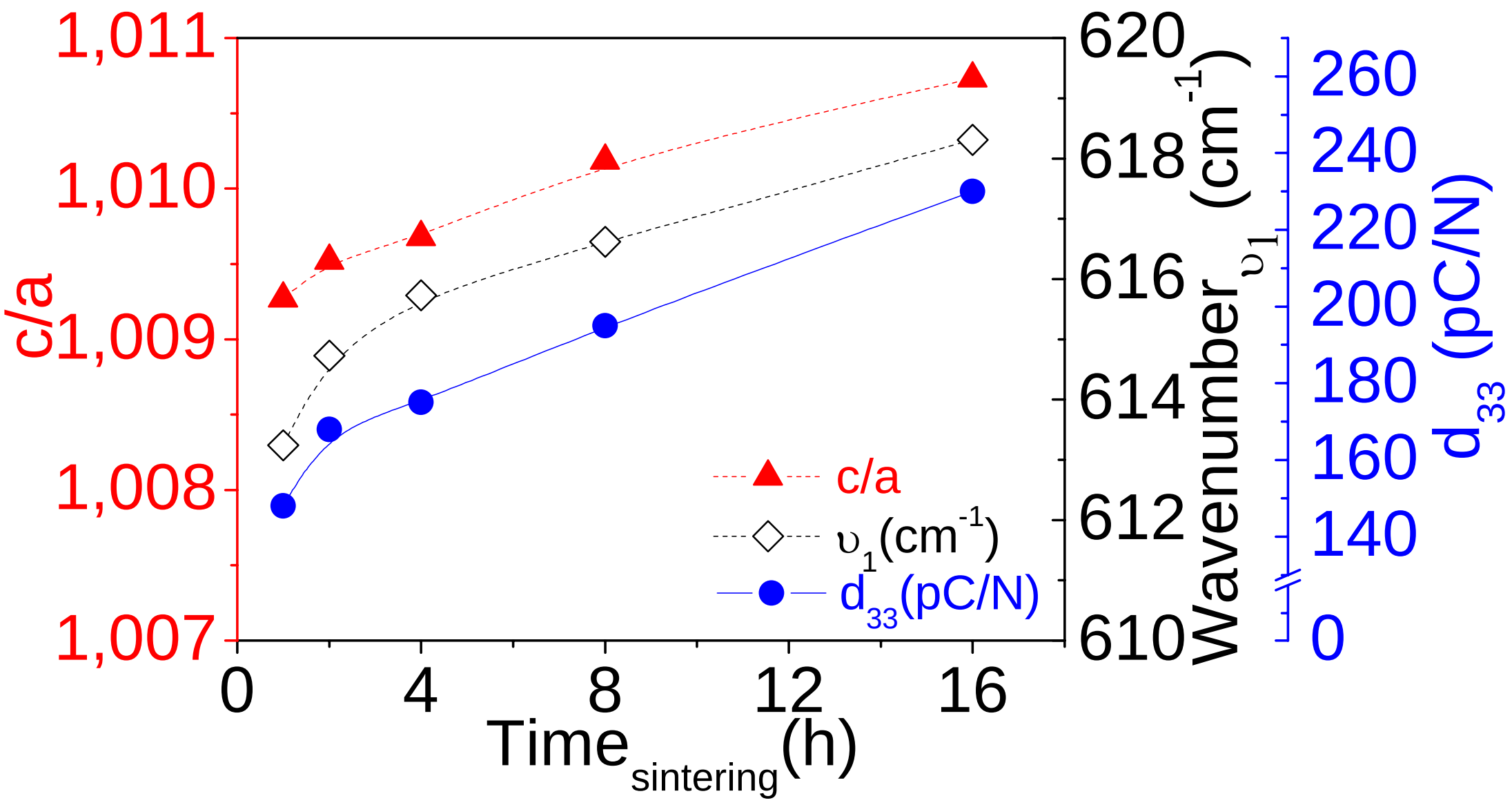

**F.Rubio-Marcos et.al. Figure. 2**

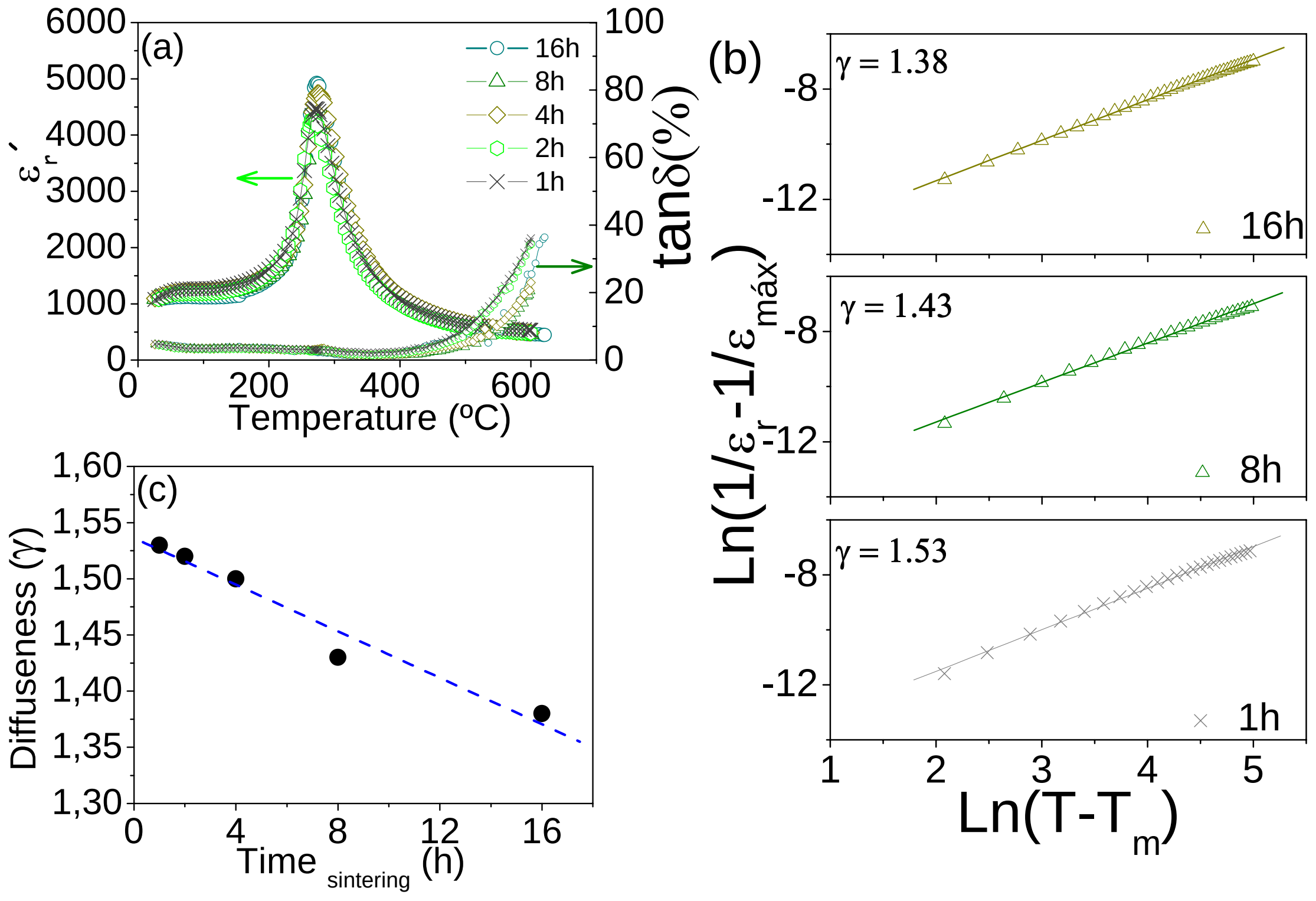

F.Rubio-Marcos et.al. Figure. 3

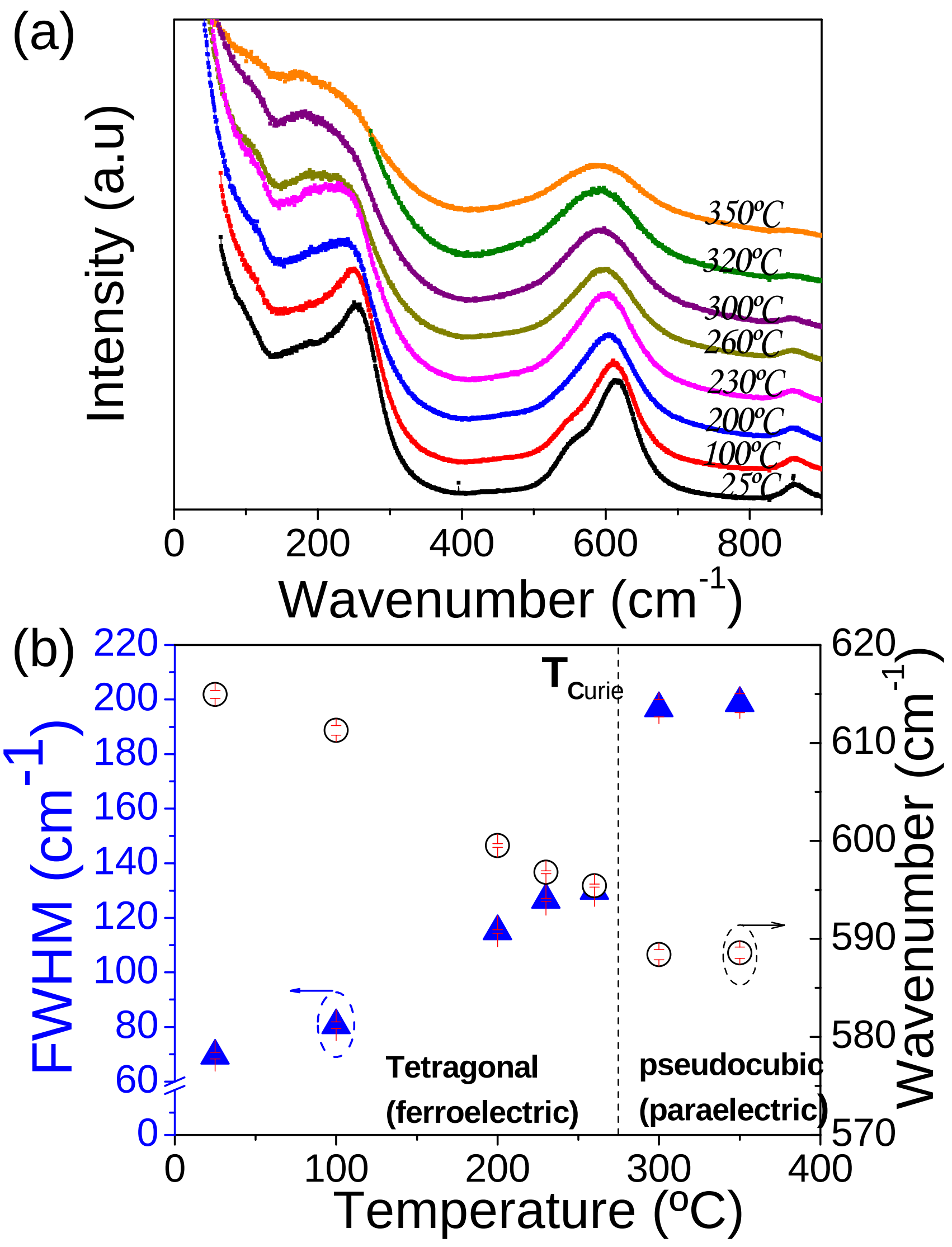

**F.Rubio-Marcos et.al. Figure. 4**